\documentclass[12pt,showpacs,showkeys,prb,preprint]{revtex4}
\usepackage{hyperref}
\usepackage{amsmath}
\usepackage{amssymb}
\usepackage{graphicx}

\begin{document}
\draft

\title{Exact solutions for reconnective annihilation in magnetic configurations with three sources}
\date{\today}
\author{E. Tassi}
\author{V.S. Titov}
\author{G. Hornig}
\affiliation{Theoretische Physik IV, Ruhr-Universit\"at Bochum, 44780 Bochum, Germany}
\email{tassi@tp4.ruhr-uni-bochum.de}

\begin{abstract}
Exact solutions of the steady resistive three dimensional (3D) magnetohydrodynamics (MHD) equations in cylindrical coordinates for an incompressible plasma are presented. The solutions are translationally invariant along one direction and in general they describe a process of reconnective annihilation in a curved current layer with non vanishing magnetic field. In the derivation of the solutions the ideal case with vanishing resistivity and electric field is considered first and then generalized to include the effects of finite electric field and resistivity. Particular attention is devoted to the analysis how the latter ones influence the presence of singularities in the solutions. In this respect comparisons with the purely two-dimensional case are made and the resulting important differences are highlighted. Finally, applications of the solutions for modeling an important class of solar flares are discussed.

\end{abstract}


\maketitle

\section{Introduction}

Magnetic reconnection is an effective mechanism for restructuring
the magnetic field and converting magnetic energy into other forms of energy in
plasmas with a high electric conductivity.
 This process plays a key role in large scale cosmic phenomena such as
solar flares and geomagnetic substorms.
 Since the pioneering works of Dungey\cite{Du53}, Parker\cite{Pa57}, Sweet\cite{Sw58} and Petschek\cite{Pe64}, where the
basic physics of magnetic reconnection was clarified, many exact
two-dimensional (2D) solutions describing some simplified versions of this
process were found\cite{So75,Cr95,Pr00,Ta02}.

 Moreover, over the last few years there was a similar progress
in the theory of three-dimensional (3D) reconnection.
 In particular, some simplified forms of this process at  null points of 3D magnetic fields have been found as well\cite{Cr96,Cr99,Me02}. 

 In this paper we present a generalization of exact solutions
for a 2D curvilinear geometry\cite{Ta02}, the so-called two and a half
dimensions (2${1 \over
2}$D) case, where a translational invariant component of velocity and magnetic
field  along the third direction exists.
 These solutions describe steady incompressible resistive magnetohydrodynamics (MHD) flows
in a configuration with non-vanishing magnetic field.
 In some respects they resemble the 2D solutions describing a
particular type of magnetic reconnection which is called reconnective
annihilation\cite{Pr00}.  So we will also use this term further on to denote the process described
by our solutions.

 It should be noted also that when the present article was in
preparation, a paper of Watson and Craig\cite{Wa02} appeared, where similar
solutions have been presented. Since we found these solutions
independently, our considerations and interpretations differ in many respects. 

In sections \ref{s:beqs} and \ref{s:fos}, respectively, the basic
equations and the form of solutions are described.  In section
\ref{sec:ideal} we derive the solutions in the limit of ideal MHD and
discuss their properties, while in section \ref{s:rs} we consider how
these solutions are modified in the case of non-vanishing resistivity.
The conclusions are drawn in section \ref{s:c}.

\section{Basic equations}
\label{s:beqs}

The set of MHD equations for an incompressible  plasma with uniform
density and resistivity consists of the equation of motion  
\begin{equation}    \label{e:mot}
({\bf v}\cdot\nabla){\bf v}=-\nabla p +{{(\nabla\times{\bf B}}})\times{\bf B},
\end{equation}
the Ohm's law 
\begin{equation}  \label{e:Ohm}
{\bf E}+{\bf v}\times{\bf B}=\eta\nabla\times{\bf B}
\end{equation}
and the divergence-free conditions for the velocity ${\bf v}$ and the
magnetic field ${\bf B}$  
\begin{equation} \label{e:divfree}
\begin{split}
&\nabla\cdot{\bf v}=0,\\
&\nabla\cdot{\bf B}=0.\\
\end{split}
\end{equation}
All equations here are written in a dimensionless form such that ${\bf
  B}$ and ${\bf v}$ are normalized to $B_e$ and
$v_{Ae}$ respectively, which are characteristic values of  the magnetic field and of
the Alfv\a'en velocity. ${\mathbf E}$ represents the dimensionless electric field while $\eta$ corresponds to the inverse magnetic Reynolds number.\\ 
The current density ${\bf j}$ is determined separately by
Amp\a`ere's law 
\begin{equation}   \label{e:Ampe}
{\bf j}=\nabla \times {\bf B}.
\end{equation}
Consider a cylindrical coordinate system $(r,\theta,z)$ where $r$ and $\theta$ 
are related to the Cartesian coordinates $(x,y)$ in the following way
\begin{equation}  \label{e:polcoord}
x=r\sin \theta, \qquad y=r \cos\theta -d,
\end{equation}
where $d>0$ so that the pole is below the plane $y=0$. 
Assume that in this coordinate system the
functions ${\mathbf B}$ and ${\mathbf v}$ depend only on $r$ and
$\theta$. Then they can be written as follows 
 
\begin{equation}   \label{e:ans}
(B_r, B_{\theta}, B_z)=\left(\frac{1}{r}{\frac{\partial A}{\partial
      \theta}}, -{\frac{\partial A}{\partial r}}, H \right), \qquad
(v_r, v_{\theta}, v_z)=\left(\frac{1}{r}{\frac{\partial \psi}{\partial
      \theta}}, -{\frac{\partial \psi}{\partial r}}, V \right), 
\end{equation}
where $A$, $H$, $\psi$ and $V$ are functions of $r$ and $\theta$ which are to be
found. In particular, for $H=V=0$ this is the usual representation of
two-dimensional magnetic and incompressible velocity fields in terms
of a flux function $A$ and a stream function $\psi$, respectively.  
 
By using this representation we obtain from Eq. (\ref{e:mot}) that the
functions $A$, $H$, $\psi$ and $V$ must satisfy the following two
equations 
\begin{eqnarray}   \label{e:mot2}
[\psi,{\nabla}^2 \psi]=[A,{\nabla}^2 A], 
\end{eqnarray}
\begin{eqnarray}   \label{e:mot1}
[V,\psi]-[H,A]=Cr, 
\end{eqnarray}
where $C$ is an arbitrary constant and the Poisson brackets are used such that
\begin{equation*}
[f,g]={\frac{\partial f}{\partial r}}{\frac{\partial g}{\partial
    \theta}}-{\frac{\partial g}{\partial r}}{\frac{\partial
    f}{\partial \theta}}. 
\end{equation*}
Similarly, Eq. (\ref{e:Ohm}) gives the following two equations 
\begin{eqnarray}    \label{e:Ohm1}
[H,\psi]+[A,V]=\eta r{\nabla}^2 H,
\end{eqnarray} 
\begin{eqnarray}    \label{e:Ohm2}
E_z r+[\psi,A]=-\eta r{\nabla}^2 A.
\end{eqnarray} 

\section{Form of the solutions}
\label{s:fos}

For the system (\ref{e:mot2})--(\ref{e:Ohm2}) we seek solutions of the form 
 
\begin{equation} \label{e:ans2}        
\begin{split}
&A(r,\theta)=A_1(r)\theta+A_0(r),\\
&\psi(r,\theta)=\psi_1(r)\theta+\psi_0(r),\\
&H(r,\theta)=H_1(r)\theta+H_0(r),\\
&V(r,\theta)=V_1(r)\theta+V_0(r).\\
\end{split}
\end{equation}
This form is a generalization of the ansatz used in Ref. \onlinecite{Ta02}  for a
two-dimensional configuration. Substituting (\ref{e:ans2}) into
(\ref{e:mot2})--(\ref{e:Ohm2}) provides four equations each of which
is a polynomial linear in $\theta$. Thus for each
equation the part of the polynomial not depending on $\theta$ and the
coefficient of $\theta$ must be separately equal to $0$. This yields
the following set of ordinary differential equations (ODEs): 
\begin{eqnarray}  \label{e:i1}
V_1{\psi_1}^{\prime}-{V_1}^{\prime}{\psi_1}-H_1{A_1}^{\prime}+{H_1}^{\prime}{A_1}=0
\end{eqnarray}
\begin{eqnarray}  \label{e:i2}
A_1{H_0}^{\prime}-{H_1}{A_0}^{\prime}+V_1{\psi_0}^{\prime}-{\psi_1}{V_0}^{\prime}=Cr
\end{eqnarray}
\begin{eqnarray}  \label{e:i3}
{{\psi_1}^{'}\over r}{\left(r{\psi_1}^{'}\right)}^{'}
-\psi_1{\left[{1\over r}{(r{\psi_1}')}^{'}\right]}^{'}=
{{A_1}'\over r}{\left({r{A_1}'}\right)}^{'}-A_1{\left[{1\over
      r}{(r{A_1}')}^{'}\right]}^{'} 
\end{eqnarray}
\begin{eqnarray}  \label{e:i4}
{\frac{{\psi_0}'}{r}}{\left({r{\psi_1}'}\right)}'-\psi_1{\left[\frac{1}{r}{(r{\psi_0
}')}'\right]}^{'}={\frac{{A_0}'}{r}}{\left({r{A_1}'}\right)}^{'}-A_1{\left[
\frac{1}{r}{( r{A_0}')}^{'}\right]}^{'}
\end{eqnarray}
\begin{eqnarray}  \label{e:i5}
{A_1}^{\prime}V_1-A_1{V_1}^{\prime}+\psi_1{H_1}^{\prime}-H_1{\psi_1}^{\prime}-\eta
r{H_1}^{\prime\prime}-\eta {H_1}^{\prime}=0 
\end{eqnarray}
\begin{eqnarray}  \label{e:i6}
V_1{A_0}^{\prime}+\psi_1{H_0}^{\prime}-A_1{V_0}^{\prime}-H_1{\psi_0}^{\prime}-\eta
r{H_0}^{\prime\prime}-\eta {H_0}^{\prime}=0 
\end{eqnarray}
\begin{eqnarray}  \label{e:i7}
  \psi_1 'A_1-\psi_1 A_1 '+\eta ({A_1}^{\prime}+ r {A_1}^{\prime\prime})=0
\end{eqnarray}
\begin{eqnarray}  \label{e:i8}
  E_z+\frac{1}{r}[\psi_0 'A_1-\psi_1 A_0 '+\eta ({A_0}^{\prime}+ r
  {A_0}^{\prime\prime})]=0. 
 \end{eqnarray}
Here the prime stands for the derivative with respect to $r$.
The above system consists of 8 ODEs for 8 unknowns of one
variable. Therefore the ansatz (\ref{e:ans2}) is compatible with the
original system of partial differential equations. In our stationary
case $\nabla \times {\bf E}=0$, which together with Eq. (\ref{e:ans2})
yields a uniform $z$-component of the electric field.

One can also notice that Eqs. (\ref{e:i3}), (\ref{e:i4}), (\ref{e:i7})
and (\ref{e:i8}) are the same as for the purely two-dimensional (2D)
case corresponding to setting $H\equiv V \equiv 0$. Thus, the 2D
equations are a limiting case of our 2$\frac{1}{2}$ D case and, what is
even more important, they are decoupled from the rest equations of the
system. Solutions of Eq. (\ref{e:i3}), (\ref{e:i4}), (\ref{e:i7}) and
(\ref{e:i8}) have already been presented in Ref. \onlinecite{Wa02} and we can
simply use these solutions for solving our more general problem. It is
worth noticing that, for known $A_1$, ${A_0}^{\prime}$,
$\psi_1$ and ${\psi_0}^{\prime}$, the system of equations
(\ref{e:i1}), (\ref{e:i2}), (\ref{e:i5}) and (\ref{e:i6}) is linear in
the functions $H_1$, ${H_0}^{\prime}$, $V_1$ and ${V_0}'$.

\section{Ideal solutions} \label{sec:ideal}

In this section Eqs. (\ref{e:i1})--(\ref{e:i8}) are analyzed in the
limit of vanishing resistivity $\eta=0$. We start by considering the
equipotential case, i.e. the case where the electric potential is
constant along the $z$ axis so that $E_z$ vanishes as well. In this
case, according to (\ref{e:Ohm2}), $[\psi,A]=0$ and (\ref{e:ans2}) the
relationship%
\begin{equation}
\psi=\alpha A, \qquad \alpha=\text{constant}
\end{equation}
is valid. Eqs. (\ref{e:i3}), (\ref{e:i4}), (\ref{e:i7}) and (\ref{e:i8}) then yield
\begin{eqnarray}    \label{e:A1p1idE0}
A_1=C_1 \ln r +C_2, \qquad \psi_1=\alpha A_1,
\end{eqnarray}
\begin{eqnarray}
A_0=C_3 r^2+C_4 \ln r,  \qquad \psi_0=\alpha A_0,
\end{eqnarray}
where $C_1$, $C_2$, $C_3$ and $C_4$ are arbitrary constants. These
solutions describe a magnetic configuration with a field-aligned flow
in the $(r,\theta)$ plane. In this configuration there are one
magnetic null point and one stagnation point, both are located at
${\bf r}^*=[r_c,-(2 C_3 {r_c}^2+C_4)/ C_1]$. Here $r_c \equiv
\exp(-C_2/C_1)$ is denoted as a critical radius.\\  
It is not difficult to see now that Eqs. (\ref{e:i1}) and (\ref{e:i6})
can be satisfied in the ideal limit by 
\begin{equation} \label{e:V1H1idE0}
V_1=c_1 A_1,  \qquad H_1=c_2 A_1,
\end{equation}
where $c_1$ and $c_2$ are other arbitrary constants.
Therefore $V_1$ and $H_1$ also vanish together with $A_1$ and $\psi_1$
at $r_c$. Then the evaluation of (\ref{e:i2}) at $r_c$ requires
 $C \equiv 0$. Substituting of (\ref{e:V1H1idE0}) into (\ref{e:i2}) and
(\ref{e:i6}) yields  
\begin{equation} \label{e:V0H0idE0}
\begin{split}
&V_0=-\frac{c_1}{{\alpha}^2-1}(A_0-\alpha
\psi_0)-\frac{c_2}{{\alpha}^2-1}(\alpha A_0-\psi_0)={c_1}(C_3 r^2 +
C_4\ln r)+C_5,\\  
&H_0=-\frac{c_1}{{\alpha}^2-1}(\alpha A_0-
\psi_0)-\frac{c_2}{{\alpha}^2-1}(A_0-\alpha \psi_0)={c_2}(C_3 r^2 +
C_4 \ln r)+C_6,\\. 
\end{split}
\end{equation}   
where $C_5$ and $C_6$ are arbitrary constants.\\
Evaluating $B_z$ and $v_z$ at ${\bf r}^*$ we obtain
\begin{equation}
\begin{split}
&\left. B_z \right\vert_{{\bf r}={\bf r}^*}=-2c_2C_3{r_c}^2\left(\ln
  r_c-\frac{1}{2}\right)-\frac{2 c_2
  C_2}{C_1}\left(C_3{r_c}^2+\frac{C_4}{2}\right)+C_6,\\ 
&\left. v_z \right\vert_{{\bf r}={\bf r}^*}=-2c_1C_3{r_c}^2\left(\ln
  r_c-\frac{1}{2}\right)-\frac{2 c_1
  C_2}{C_1}\left(C_3{r_c}^2+\frac{C_4}{2}\right)+C_5,\\ 
\end{split}
\end{equation} 
which in general do not vanish. Therefore, contrary to the 2D case, in
2$\frac{1}{2}$D we generally have neither
nulls nor stagnation points in the limit $E_z=\eta=0$.   

\begin{figure}
\begin{center}
\includegraphics[width=12cm]{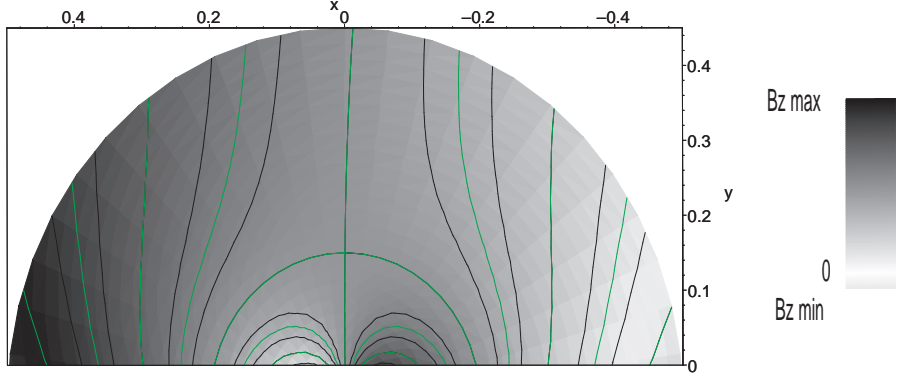}
\end{center}
\caption{Magnetic (solid) and velocity (dashed) field lines
  superimposed on the distribution of the $z$ component of the
  magnetic field shown in gray half-tones. The parameters used in the
  plot are $E_z=\eta=0$, $r_c=0.2$, $\alpha=2$, $C_1=1.2$, $C_3=0.2$,
  $C_4=-0.08 C_3$, $C_2=-C_1 \ln 0.2$, $C_6=1$, $c_2=0.8$, $d=0.05$.}  
\label{fig:fig1}
\end{figure}

The velocity and magnetic field lines are presented in
Fig. \ref{fig:fig1}. They are superimposed on the distribution of the
magnetic $z$ component. The poloidal components ($r, \theta$) and the
nonuniform parts of the toroidal components ($z$) of our ${\bf v}$ and
${\bf B}$ are proportional to each other but generally with different
coefficients of proportionality. This means that in our case the flow
reshuffles the magnetic field lines inside each of the magnetic
surfaces $A(r, \theta)=\text{constant}$ but it preserves the surfaces
themselves. It can be noticed also that the distribution of magnetic
flux can be imagined as generated by three sources with alternating
polarities lying on the plane $y=0$.\\ 
Consider now how the above ideal equipotential solution is modified in
the presence of a finite $z$ component of the electric field. In two
dimensions it has been shown in Ref. \onlinecite{Ta02} that a non-vanishing $E_z$
causes the appearance of a singularity at $r=r_c$ in the current
density, vorticity and azimuthal components of the magnetic and
velocity fields. Indeed, one class of solutions can be written as
follows  
\begin{eqnarray}    \label{e:A1p1id}
A_1=C_1 \ln r +C_2, \qquad \psi_1=\alpha A_1,
\end{eqnarray}
\begin{eqnarray}    \label{e:A0p0id}
{A_0}^{\prime}=\frac{\alpha}{{{\alpha}^2-1}}{\frac{E_z
    r}{{A_1}}}+\frac{a}{\alpha}r+\frac{b}{{\alpha r}}, \qquad
{\psi_0}^{\prime}=\frac{1}{{{\alpha}^2-1}}{\frac{E_z
    r}{{A_1}}}+ar+\frac{b}{r}, 
 \end{eqnarray}
where $a$ and $b$ are arbitrary constants. From here one can see that
the magnetic flux piles up at the separatrix $r=r_c$ to produce there
the above mentioned singularity. Coming back to our  2$\frac{1}{2}$ D
problem, we notice that Eqs. (\ref{e:i1}),
(\ref{e:i2}),(\ref{e:i5}) and (\ref{e:i6}) do not depend on $E_z$.
Therefore the expressions (\ref{e:V0H0idE0}) are still applicable for
$E_z \neq 0$, if one uses for ${A_0}^{\prime}$ and ${\psi_0}^{\prime}$
the expressions (\ref{e:A0p0id}). The solutions for $V_1$, $H_1$,
${V_0}^{\prime}$ and ${H_0}^{\prime}$ are then given by  
\begin{eqnarray} \label{e:V1H1id}
V_1=c_1 A_1,  \qquad H_1=c_2 A_1,
\end{eqnarray}
\begin{eqnarray}
{H_0}^{\prime}=-\frac{c_1}{{\alpha}^2-1}\frac{E_z
  r}{A_1}+\frac{c_2}{\alpha}\left(ar+\frac{b}{r}\right), \qquad
{V_0}^{\prime}=\frac{c_1}{\alpha}\left(ar+\frac{b}{r}\right)-\frac{c_2}{{\alpha}^2-1}
\frac{E_z r}{A_1}. 
\end{eqnarray}
These expressions show that, as in the 2D case, the presence of a
non-vanishing $E_z$ leads to the appearance of singularities in the
distributions of physical values.   

\begin{figure}
\includegraphics[width=10cm]{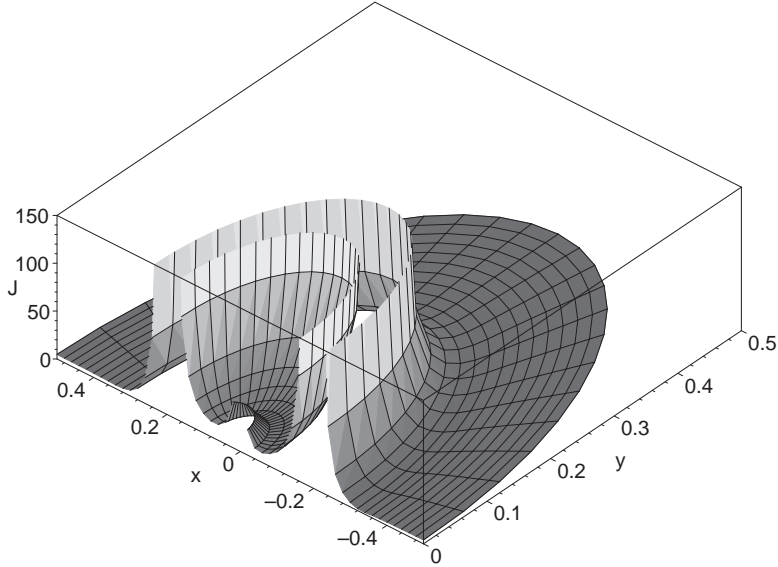}
\caption{Distribution of the current density $J \equiv |{\bf j}|$ in the case
  of $\eta=0$ and finite $E_z$. The other used parameters in the plots
  are $c_1=0.5$, $c_2=0.4$, $r_c=0.2$, $a=0$, $b=1$, $\alpha=8/9$,
  $C_1=-0.9/\ln 0.2$, $C_2=1$, $d=0.05$.} 
\label{fig:picture2}
\end{figure}

In particular, there is a singularity in the current density
distribution (Fig. \ref{fig:picture2}) as well as in the distributions
of ${A_0}^{\prime}$, ${\psi_0}^{\prime}$, ${H_0}^{\prime}$ and
${V_0}^{\prime}$. This means in turn that both the toroidal ($z$) and
poloidal components of the current density become singular, which is
also clear from their explicit expressions: 
\begin{equation}
\begin{split}
& j_r=\frac{H_1}{r},\\
& j_{\theta}=-H_1'\theta-H_0',\\
& j_z=-A_0'-rA_0''.\\
\end{split}
\end{equation}
In other words, these 2${1 \over 2}$D solutions inherit the
singularities from the corresponding 2D solutions although the
interpretation of the inherited singularities compared 
to the original ones differs in several important respects. 
 
 As was shown above, the equipotential 2D magnetic configuration has a
null point or, if one extends the system to three dimensions, a null
line parallel to the invariant $z$ direction.
 If an electric field ${\bf E}$ appears in the system, it may have
just one $z$ component due to the assumed two-dimensionality of the
flow.  
 The steadiness of the flow requires also that this electric field
must be uniform. 
 The latter implies in particular the presence of the electric field
at the place where the null point (or null line) was initially located
in the equipotential configuration. 
 According to the frozen-in law condition, however, the finite value
of ${\bf E}$ is sustained by the inductive field $-{\bf
  v}\times {\bf B}$ only.
 So, kinematically, the presence of ${\bf E}\ne {\bf 0}$ at a magnetic
null point would compel the velocity ${\bf v}$ to be infinite there. 
 In our self-consistent approach, incorporating both kinematics and
dynamics of plasma, any velocity singularity may coexist only with an
appropriate magnetic singularity, appearing at the same point in the
result of the corresponding force balance. 
 This is the reason why the magnetic null point of the equipotential
configuration transforms into the proper magnetic singularity when
passing in the considered family of exact 2D solutions to the
configuration with a non-vanishing electric field. 

The situation becomes different for our 2${1 \over 2}$D
configurations, which generally have no longer null points. 
 Therefore we cannot appeal to the above ``null-point'' argument
to explain the origin of the inherited singularities.
 The desirable explanation in fact can be found considering 
the consequences of the presence of field lines with a finite
longitudinal voltage drop \cite{Sc88}, denoted also as {\it singular
  magnetic field lines} \cite{Pr89}. Our 2${1 \over 2}$D equipotential
configuration has in place of the 
null line a straight magnetic field line parallel to $z$ axis. 
 Passing to non-equipotential configurations in our family of
solutions leads to the appearance of a constant $z$ component of the
electric field, which inevitably creates the above mentioned voltage drop along
a straight magnetic field line. 
 Kinematically, such a voltage drop in a plasma with an infinite
conductivity may be sustained by a suitable velocity singularity only
\cite{Pr89}.
 Our self-consistent consideration, incorporating plasma dynamics,
shows that such a velocity singularity gives birth to an appropriate
magnetic singularity by analogy with the 2D case. 
 Thus, our stationary 2${1 \over 2}$D ideal MHD solutions provide a
particular but explicit realization of singular magnetic field lines,
whose properties were kinematically described first by Schindler et
al. \cite{Sc88} in a more general non-stationary case, where the voltage drop
was localized at a finite part of such lines. 

 The above qualitative consideration shows that the appearance of
singularities at the null line or singular magnetic field line can be
anticipated if one combines the results of an analysis based on both the
kinematics and the dynamics of a plasma. 
 However, our explicit solutions reveal a much less obvious feature of
this process, namely, that the singularity appears not only at the
null or singular magnetic field lines but also at the whole magnetic
separatrix surface $r=r_{\rm c}$ containing such lines.    
 In a more simple neutral X-type point configuration a similar fact
follows from the frozen-in law and incompressibility conditions if one
assumes also that the resulting plasma flow crosses one of the two
 separatrices and is parallel to the other\cite{Pr94}. 
 These conditions are sufficient for the appearance of a singularity
along the separatrix which is not crossed by the flow. 
 The self-consistent incorporation of the plasma dynamics just makes
the type of such a singularity more precise. 
 One can also see this from the ideal MHD limit of Craig-Henton solution
\cite{Cr95} and from the same limit of more general solutions \cite{Pr00}. 
 It is not difficult to show that the considered point of view remains
valid for our 2D and 2${1 \over 2}$D solutions in curvilinear geometry
as well. 

 Thus, the above discussion suggests that the ``extension" of the
singularity from the null or singular field line to the whole
separatrix surface is somehow effected by the incompressibility
condition. 
 This suggestion, however, looks not convincing enough if one
remembers that in the three-dimensional case the velocity singularity
at the fan separatrix surfaces of magnetic nulls appears already in a
purely kinematic approach \cite{Pr96} without any involvement
of the incompressibility condition. 
 So the complete clarification of the nature of the separatrix
singularities is still a matter of the future development of the
theory. 
 Irrespective of the ultimate answer to this question, it is important
to study how such singularities are resolved in the framework of a
self-consistent MHD approach by a finite resistivity, which is an
issue of the next section.

\section{Resistive solutions}
\label{s:rs}

Let us now consider the complete system
(\ref{e:i1})--(\ref{e:i8}). For the functions $A_1$, $\psi_1$, $A_0'$
and $\psi_0'$ we can use the following solutions  
\begin{eqnarray} \label{e:A1p1res}
A_1=C_1 \ln r +C_2, \qquad \psi_1=\alpha A_1,
\end{eqnarray}
\begin{eqnarray}  \label{e:A0p0res}
{A_0}^{\prime}(r)={-{{\exp\left[-{({\alpha}^2-1){(C_1\ln r
            +C_2)}^2}\over{2\eta\alpha}\right]\over{\eta r}}}}
\int_{r_c}^r{\frac{{E_z
      t-A_1{{1-{\alpha}^2}\over{{\alpha}^2}}\left({at+{b/t}}\right)dt}}{{{\exp 
\left[-{({\alpha}^2-1){(C_1\ln t +C_2)}^2}\over{2\eta\alpha}\right]}}}},  
\end{eqnarray}
\begin{eqnarray}
{\psi_0}^{\prime}(r)={1\over
  {\alpha}}\left[{A_0}^{\prime}(r)-{{1-{\alpha}^2}\over{\alpha}}\left(
    ar+{b\over r}\right)\right] 
\end{eqnarray}
derived in Ref. \onlinecite{Wa02}. These solutions describe a 2D reconnective
annihilation in a curved current layer formed on one of the magnetic
separatrices when the other is crossed by a sheared flow. The magnetic
and velocity fields have a null point and a stagnation point,
respectively, whose positions, contrary to the case considered in
Sec. \ref{sec:ideal}, are in general not coincident. Considering now
the 2${1 \over 2}$D problem, we first notice that
Eqs. (\ref{e:V1H1id}) are solutions of the system for a finite
resistivity as well. Therefore the problem is reduced to finding solutions for
$H_0'$ and $V_0'$. By means of Eq. (\ref{e:i2}) we can express $V_0'$
in the form 
\begin{equation}
V_0'=\frac{A_1 H_0'-H_1 A_0'+V_1 \psi_0'}{\psi_1}   
\end{equation}
which, with the help of (\ref{e:A1p1res}) and (\ref{e:V1H1id}), can be reduced to
\begin{equation}
V_0'=\frac{1}{\alpha}({H_0'}-{c_2}A_0'+{c_1}\psi_0').
\end{equation}
Inserting this expression into Eq. (\ref{e:i6}) and again using
(\ref{e:A1p1res}) and (\ref{e:V1H1id}), we obtain the following
equation 
\begin{equation}
\eta r
H_0''+\left[\frac{(1-{\alpha}^2)}{\alpha}A_1+\eta\right]H_0'+A_1\left[\left(
\frac{c_1}{\alpha}+c_2\right)\psi_0'-\left(\frac{c_2}{\alpha}+c_1
  \right)A_0' \right]=0. 
\end{equation}
This linear equation can be solved by the method of the variation of
parameters with the following result: 
\begin{equation} \label{e:H0sol1}
\begin{split}
&{H_0}^{\prime}=\left[\frac{1}{{\alpha}\eta r}\int ds
  {\left(V_1(\alpha
      {A_0}^{\prime}-{\psi_0}^{\prime})+H_1({A_0}^{\prime}-\alpha{\psi_0}^{\prime}) 
\right)}  \right. \\    
&\left. \qquad \exp\left(-\frac{({\alpha}^2-1)}{\eta
      \alpha}\int\frac{A_1}{t}dt\right)+c_5\right]\exp\left(\frac{({\alpha}^2-1)}{\eta
 \alpha}\int\frac{A_1}{s}ds\right). \\ 
\end{split} 
\end{equation}
By means of (\ref{e:V1H1id}), (\ref{e:A1p1res}) and (\ref{e:A0p0res})
the solution (\ref{e:H0sol1}) can be written as follows 
\begin{equation}
\begin{split}
&{H_0}^{\prime}={\exp\left[{-({\alpha}^2-1){(C_1\ln r
        +C_2)}^2}\over{2\eta\alpha}\right]}\left[\frac{1}{{\alpha}\eta
    r} \int ds
  \left(c_1({\alpha}^2-1)\left(\psi_0'-as-\frac{b}{s}\right)\right.\right.\\ 
&\left.\left. 
-c_2\frac{{\alpha}^2-1}{\alpha}\left(as+\frac{b}{s}\right)\right)(C_1
\ln s+C_2){ { \exp \left({({\alpha}^2-1){(C_1\ln s +C_2)}^2}
\over{2\eta\alpha}\right)}}+c_5\right].\\  
&
\end{split}  
\end{equation}

\begin{figure} 
\includegraphics[width=14cm]{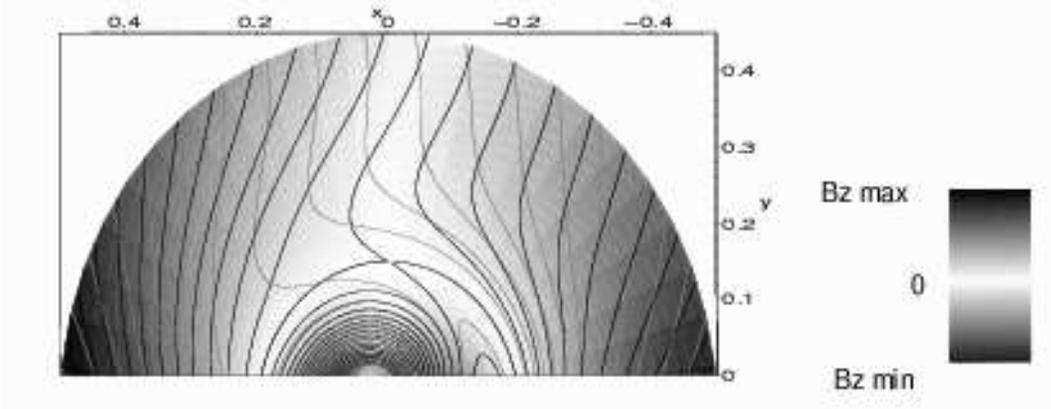}

\caption{Integral lines of the poloidal magnetic (in black) and
  velocity (in gray) fields, superimposed on the distribution of the
  $z$ component of the magnetic field shown in gray half-tones. The
  parameters used for the plot are $\eta=10^{-2}$, $E_z=0.5$,
  $c_1=0.5$, $c_2=0.4$, $r_c=0.2$, $a=0$, $b=1$, $\alpha=8/9$,
  $C_1=-0.9/\ln 0.2$, $C_2=1$, $d=0.05$.} 
\label{fig:picture3}   
\end{figure}

The resulting magnetic field configuration is represented in
Fig. \ref{fig:picture3}. The plot refers to the particular case where
the line $(r=r_c, \theta=0)$ is a magnetic null line. A more general
configuration without this null line may be obtained by simply adding
any constant value to the corresponding $B_z$ distribution. This would
give us an example of the resistive solution which in the limit of the
ideal MHD describes the above mentioned configuration with singular
magnetic field line.

\begin{figure}
\includegraphics[width=10cm]{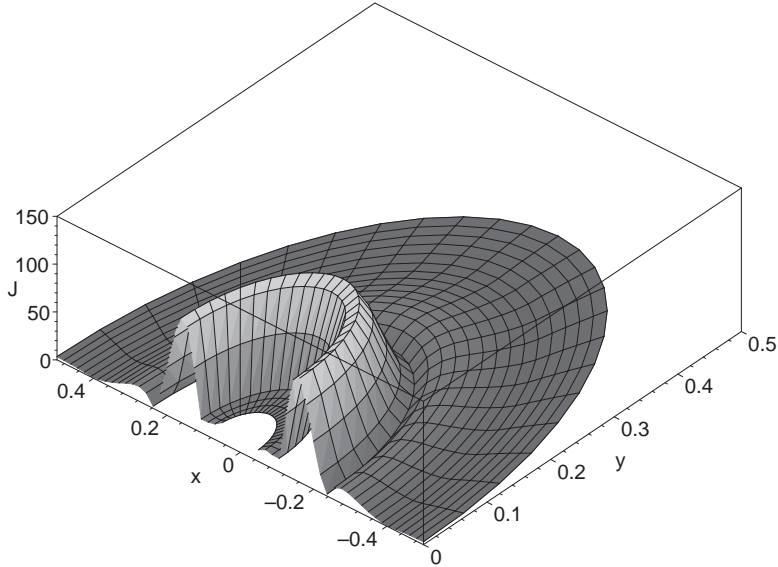}
\caption{Distribution of current density for the resistive case. 
The parameters used for this plot are the same used for the plot in
fig. \ref{fig:picture3}.}
\label{fig:picture4}
\end{figure}

In Fig. \ref{fig:picture4} the absolute value of the current density
is plotted. Comparing this plot with the one shown in
Fig. \ref{fig:picture2}, one can see the effect of introducing a
finite resistivity, which indeed resolves the singularity at the
critical radius.\\ 
As already noticed in Ref. \onlinecite{Ta02}, the configuration of the magnetic
field in the $xy$ plane described by our resistive solutions is of
particular interest for modeling the reconnection process in a special
large class of solar flares. One can see  from Fig. \ref{fig:picture3}
that the magnetic field here is generated by three sources with
alternating polarities on the plane $y=0$. Observations \cite{Ni97}
show that a large fraction of solar flares occurs namely in
configurations with three photospheric magnetic sources. The
2$\frac{1}{2}$D solutions presented in this paper provide a
generalization which makes the previous 2D model more realistic.   

\section{Conclusions}
\label{s:c}
We presented a class of steady solutions of resistive incompressible MHD
equations in cylindrical coordinates.
 These solutions describe three-dimensional velocity and magnetic
fields with a translational invariance along the $z$ axis in a
half-space corresponding to the solar corona.
 Both fields are represented as a superposition of poloidal and
toroidal ($z$) components, so that the resulting fields have no null
points in the volume.
 The poloidal fields, however, have one null line along which two
separatrix surfaces intersect.
 One of the separatrices is the same for poloidal velocity and
magnetic fields and it is a segment of cylinder with the edges at the
photospheric boundary.
 The other separatrices are different and they intersect the
cylindrical separatrix along the null lines of poloidal fields in
such a way that these lines lie at some distance from each other.
 The resulting plasma flow intersects the non-cylindrical
separatrix of the poloidal magnetic field to produce a strong shearing
motion along the cylindrical separatrix, where a strong current layer
is formed.
 Thus, our solution is a curvilinear analog of the planar
reconnective magnetic annihilation considered earlier in Ref. \onlinecite{Cr95}
and generalized in Ref. \onlinecite{Pr00}.
 In addition to the curvature it has two more features of interest.
 First, there are non-vanishing and non-uniform toroidal components of the
velocity and magnetic field in the corresponding
configuration.
 Secondly, the resulting distribution of magnetic fields on the
photosphere represents three areas of alternating polarities. 
 This feature is very interesting for modeling a wide class of solar flares which have  three photospheric magnetic sources of different polarities.  

\begin{acknowledgments}
The authors would like to gratefully acknowledge the financial support from the Volkswagen-Foundation and from the E.U. Research Training Network HPRN-CT-2000-00153.
\end{acknowledgments}

\bibliography{tassi}

\begin{thebibliography}{17}
\expandafter\ifx\csname natexlab\endcsname\relax\def\natexlab#1{#1}\fi
\expandafter\ifx\csname bibnamefont\endcsname\relax
  \def\bibnamefont#1{#1}\fi
\expandafter\ifx\csname bibfnamefont\endcsname\relax
  \def\bibfnamefont#1{#1}\fi
\expandafter\ifx\csname citenamefont\endcsname\relax
  \def\citenamefont#1{#1}\fi
\expandafter\ifx\csname url\endcsname\relax
  \def\url#1{\texttt{#1}}\fi
\expandafter\ifx\csname urlprefix\endcsname\relax\def\urlprefix{URL }\fi
\providecommand{\bibinfo}[2]{#2}
\providecommand{\eprint}[2][]{\url{#2}}

\bibitem[{\citenamefont{Dungey}(1953)}]{Du53}
\bibinfo{author}{\bibfnamefont{J.}~\bibnamefont{Dungey}},
  \bibinfo{journal}{Philos. Mag.} \textbf{\bibinfo{volume}{7}},
  \bibinfo{pages}{725} (\bibinfo{year}{1953}).

\bibitem[{\citenamefont{Parker}(1957)}]{Pa57}
\bibinfo{author}{\bibfnamefont{E.}~\bibnamefont{Parker}}, \bibinfo{journal}{J.
  Geophys. Res.} \textbf{\bibinfo{volume}{62}}, \bibinfo{pages}{509}
  (\bibinfo{year}{1957}).

\bibitem[{\citenamefont{Sweet}(1958)}]{Sw58}
\bibinfo{author}{\bibfnamefont{P.}~\bibnamefont{Sweet}},
  \emph{\bibinfo{title}{Electromagnetic Phenomena in Cosmical Physics,
  International Astronomical Union (IAU) Symp.}}, vol.~\bibinfo{volume}{6}
  (\bibinfo{publisher}{ed B. Lehnert Cambridge Univ. Press},
  \bibinfo{address}{London}, \bibinfo{year}{1958}).

\bibitem[{\citenamefont{Petschek}(1964)}]{Pe64}
\bibinfo{author}{\bibfnamefont{H.~E.} \bibnamefont{Petschek}},
  \bibinfo{journal}{Physics of solar flares, NASA Report SP-50, Washington DC}
  pp. \bibinfo{pages}{425--439} (\bibinfo{year}{1964}).

\bibitem[{\citenamefont{Sonnerup and Priest}(1975)}]{So75}
\bibinfo{author}{\bibfnamefont{B.~U.~O.} \bibnamefont{Sonnerup}}
  \bibnamefont{and} \bibinfo{author}{\bibfnamefont{E.~R.}
  \bibnamefont{Priest}}, \bibinfo{journal}{J. Plasma Phys.}
  \textbf{\bibinfo{volume}{14}}, \bibinfo{pages}{283} (\bibinfo{year}{1975}).

\bibitem[{\citenamefont{Craig and Henton}(1995)}]{Cr95}
\bibinfo{author}{\bibfnamefont{I.~J.~D.} \bibnamefont{Craig}} \bibnamefont{and}
  \bibinfo{author}{\bibfnamefont{S.~M.} \bibnamefont{Henton}},
  \bibinfo{journal}{Astrophys. J.} \textbf{\bibinfo{volume}{450}},
  \bibinfo{pages}{280} (\bibinfo{year}{1995}).

\bibitem[{\citenamefont{Priest et~al.}(2000)\citenamefont{Priest, Titov,
  Grundy, and Hood}}]{Pr00}
\bibinfo{author}{\bibfnamefont{E.~R.} \bibnamefont{Priest}},
  \bibinfo{author}{\bibfnamefont{V.~S.} \bibnamefont{Titov}},
  \bibinfo{author}{\bibfnamefont{R.~E.} \bibnamefont{Grundy}},
  \bibnamefont{and} \bibinfo{author}{\bibfnamefont{A.~W.} \bibnamefont{Hood}},
  \bibinfo{journal}{Proc. R. Soc. Lond. A} \textbf{\bibinfo{volume}{456}},
  \bibinfo{pages}{1821} (\bibinfo{year}{2000}).

\bibitem[{\citenamefont{Tassi et~al.}(2002)\citenamefont{Tassi, Titov, and
  Hornig}}]{Ta02}
\bibinfo{author}{\bibfnamefont{E.}~\bibnamefont{Tassi}},
  \bibinfo{author}{\bibfnamefont{V.~S.} \bibnamefont{Titov}}, \bibnamefont{and}
  \bibinfo{author}{\bibfnamefont{G.}~\bibnamefont{Hornig}},
  \bibinfo{journal}{Phys. Lett. A} \textbf{\bibinfo{volume}{302/5-6}},
  \bibinfo{pages}{313} (\bibinfo{year}{2002}).

\bibitem[{\citenamefont{Craig and Fabling}(1996)}]{Cr96}
\bibinfo{author}{\bibfnamefont{I.~J.~D.} \bibnamefont{Craig}} \bibnamefont{and}
  \bibinfo{author}{\bibfnamefont{R.~B.} \bibnamefont{Fabling}},
  \bibinfo{journal}{Astrophys. J.} \textbf{\bibinfo{volume}{462}},
  \bibinfo{pages}{969} (\bibinfo{year}{1996}).

\bibitem[{\citenamefont{Craig et~al.}(1999)\citenamefont{Craig, Fabling,
  Heerikhuisen, and Watson}}]{Cr99}
\bibinfo{author}{\bibfnamefont{I.~J.~D.} \bibnamefont{Craig}},
  \bibinfo{author}{\bibfnamefont{R.~B.} \bibnamefont{Fabling}},
  \bibinfo{author}{\bibfnamefont{J.}~\bibnamefont{Heerikhuisen}},
  \bibnamefont{and} \bibinfo{author}{\bibfnamefont{P.~G.}
  \bibnamefont{Watson}}, \bibinfo{journal}{Astrophys. J.}
  \textbf{\bibinfo{volume}{523}}, \bibinfo{pages}{838} (\bibinfo{year}{1999}).

\bibitem[{\citenamefont{Mellor et~al.}(2002)\citenamefont{Mellor, Priest, and
  Titov}}]{Me02}
\bibinfo{author}{\bibfnamefont{C.}~\bibnamefont{Mellor}},
  \bibinfo{author}{\bibfnamefont{E.~R.} \bibnamefont{Priest}},
  \bibnamefont{and} \bibinfo{author}{\bibfnamefont{V.~S.} \bibnamefont{Titov}},
  \bibinfo{journal}{Geophys. Astrophys. Fluid Dynamics}
  \textbf{\bibinfo{volume}{96}}, \bibinfo{pages}{153} (\bibinfo{year}{2002}).

\bibitem[{\citenamefont{Watson and Craig}(2002)}]{Wa02}
\bibinfo{author}{\bibfnamefont{P.~G.} \bibnamefont{Watson}} \bibnamefont{and}
  \bibinfo{author}{\bibfnamefont{I.~J.~D.} \bibnamefont{Craig}},
  \bibinfo{journal}{Sol. Phys.} \textbf{\bibinfo{volume}{207}},
  \bibinfo{pages}{337} (\bibinfo{year}{2002}).

\bibitem[{\citenamefont{Schindler et~al.}(1988)\citenamefont{Schindler, Hesse,
  and Birn}}]{Sc88}
\bibinfo{author}{\bibfnamefont{K.}~\bibnamefont{Schindler}},
  \bibinfo{author}{\bibfnamefont{H.}~\bibnamefont{Hesse}}, \bibnamefont{and}
  \bibinfo{author}{\bibfnamefont{J.}~\bibnamefont{Birn}}, \bibinfo{journal}{J.
  Geophys. Res.} \textbf{\bibinfo{volume}{93 A6}}, \bibinfo{pages}{5547}
  (\bibinfo{year}{1988}).

\bibitem[{\citenamefont{Priest and Forbes}(1989)}]{Pr89}
\bibinfo{author}{\bibfnamefont{E.~R.} \bibnamefont{Priest}} \bibnamefont{and}
  \bibinfo{author}{\bibfnamefont{T.~G.} \bibnamefont{Forbes}},
  \bibinfo{journal}{Sol. Phys.} \textbf{\bibinfo{volume}{119}},
  \bibinfo{pages}{211} (\bibinfo{year}{1989}).

\bibitem[{\citenamefont{Priest et~al.}(1994)\citenamefont{Priest, Titov,
  Vekstein, and Rickard}}]{Pr94}
\bibinfo{author}{\bibfnamefont{E.~R.} \bibnamefont{Priest}},
  \bibinfo{author}{\bibfnamefont{V.~S.} \bibnamefont{Titov}},
  \bibinfo{author}{\bibfnamefont{G.~E.} \bibnamefont{Vekstein}},
  \bibnamefont{and} \bibinfo{author}{\bibfnamefont{G.~J.}
  \bibnamefont{Rickard}}, \bibinfo{journal}{J. Geophys. Res.}
  \textbf{\bibinfo{volume}{99}}, \bibinfo{pages}{21467} (\bibinfo{year}{1994}).

\bibitem[{\citenamefont{Priest and Titov}(1996)}]{Pr96}
\bibinfo{author}{\bibfnamefont{E.~R.} \bibnamefont{Priest}} \bibnamefont{and}
  \bibinfo{author}{\bibfnamefont{V.~S.} \bibnamefont{Titov}},
  \bibinfo{journal}{Phil. Trans. Roy. Soc. Lond. A}
  \textbf{\bibinfo{volume}{354}}, \bibinfo{pages}{2951} (\bibinfo{year}{1996}).

\bibitem[{\citenamefont{Nishio et~al.}(1997)\citenamefont{Nishio, Yaji, Kosugi,
  Nakajima, and Sakurai}}]{Ni97}
\bibinfo{author}{\bibfnamefont{M.}~\bibnamefont{Nishio}},
  \bibinfo{author}{\bibfnamefont{K.}~\bibnamefont{Yaji}},
  \bibinfo{author}{\bibfnamefont{T.}~\bibnamefont{Kosugi}},
  \bibinfo{author}{\bibfnamefont{H.}~\bibnamefont{Nakajima}}, \bibnamefont{and}
  \bibinfo{author}{\bibfnamefont{T.}~\bibnamefont{Sakurai}},
  \bibinfo{journal}{Astrophys. J.} \textbf{\bibinfo{volume}{489}},
  \bibinfo{pages}{976} (\bibinfo{year}{1997}).

\end{thebibliography}

\end{document}